\documentclass[fleqn,12pt,twoside]{article}
\usepackage{espcrc1}
\usepackage{times}

\usepackage{graphicx}
\usepackage[figuresright]{rotating}

\newcommand{\AmS}{{\protect\the\textfont2
  A\kern-.1667em\lower.5ex\hbox{M}\kern-.125emS}}
\title{ Parity doublets in the baryon spectrum.}
\author{{ L. Ya. Glozman}\vskip5mm
  Institute for Theoretical
Physics, University of Graz, Universit\"atsplatz 5, A-8010
Graz, Austria}
\begin{document}
\maketitle

\begin{abstract}
Physics of the low-lying and high-lying hadrons in the light
flavor sector is reviewed. While the low-lying hadrons are
strongly affected by both $U(1)_A$ and spontaneous $SU(2)_L \times
SU(2)_R$ breakings, in the high-lying hadrons these symmetries are
restored. A manifestation is a persistence of the chiral multiplet
structure in both baryon and meson spectra. A fundamental origin
of this phenomenon is that effects of quantum fluctuations of both
quark and gluon fields must vanish at large $n$ or $J$ and a
semiclassical description becomes adequate. A relation between the
chiral symmetry restoration and the string picture of excited
hadrons is discussed.
\end{abstract}

\section{Introduction}
If one neglects the tiny masses of the $u$ and $d$ quarks, which
are much smaller than $\Lambda_{QCD}$, then the QCD Lagrangian
exhibits the

\begin{equation}
U(2)_L\times U(2)_R = SU(2)_L\times SU(2)_R\times U(1)_V\times U(1)_A
\label{sym}
\end{equation}

\noindent symmetry. This is because the quark-gluon interaction
Lagrangian in the chiral limit does not mix the left- and
right-handed components of quarks and hence the total QCD
Lagrangian for the two-flavor QCD can be split into left-handed
and right-handed parts which do not communicate to each other. We
know that the $U(1)_A$ symmetry of the classical QCD Lagrangian is
absent at the quantum level because of the $U(1)_A$ anomaly, which
is an effect of quantum fluctuations \cite{ANOMALY}. We also know
that the chiral $SU(2)_L\times SU(2)_R$ symmetry is spontaneously
(dynamically) broken in the QCD vacuum \cite{NJL}. That this is so
is directly evidenced by the nonzero value of the quark
condensate, $\langle \bar q q \rangle = \langle \bar q_L q_R +
\bar q_R q_L \rangle \simeq -(240 \pm 10 MeV )^3$, which
represents an order parameter for spontaneous chiral symmetry
breaking. This quark condensate  shows that in the QCD vacuum the
left-handed quarks are correlated with the right-handed antiquarks
(and vice versa) and hence the QCD vacuum breaks the chiral
symmetry. This spontaneous (dynamical) breaking of chiral symmetry
is  a pure quantum effect based upon quantum fluctuations.  As a
consequence we do not observe any chiral or $U(1)_A$ multiplets
low in the hadron spectrum.

The upper part of both baryon \cite{G1,CG1} and meson \cite{G2}
spectra almost systematically exhibits multiplets of the chiral
and $U(1)_A$ groups (for a pedagogical overview see \cite{G3}),
though a careful experimental exploration of high-lying spectra
must be done for a final conclusion. This phenomenon is referred
to as effective chiral symmetry restoration or chiral symmetry
restoration of the second kind. This is illustrated in Fig. 1,
where the excitation spectrum of the nucleon from the PDG
compilation as well as the excitation spectrum of $\pi$ and $f_0$
(with the $\bar n n = \frac{\bar u u + \bar d d}{\sqrt 2}$
content) mesons \cite{G2} are shown. Starting from the 1.7 GeV
region the nucleon (and delta) spectra show obvious signs of
parity doubling. There are a couple of examples where chiral
partners of highly excited states have not yet been seen. Their
experimental discovery would be an important task. Similarly, in
the chirally restored regime $\pi$ and $\bar n n$ $f_0$ states
must be systematically degenerate.

\begin{figure}

\begin{center}
\includegraphics*[width=5cm,angle=-90]{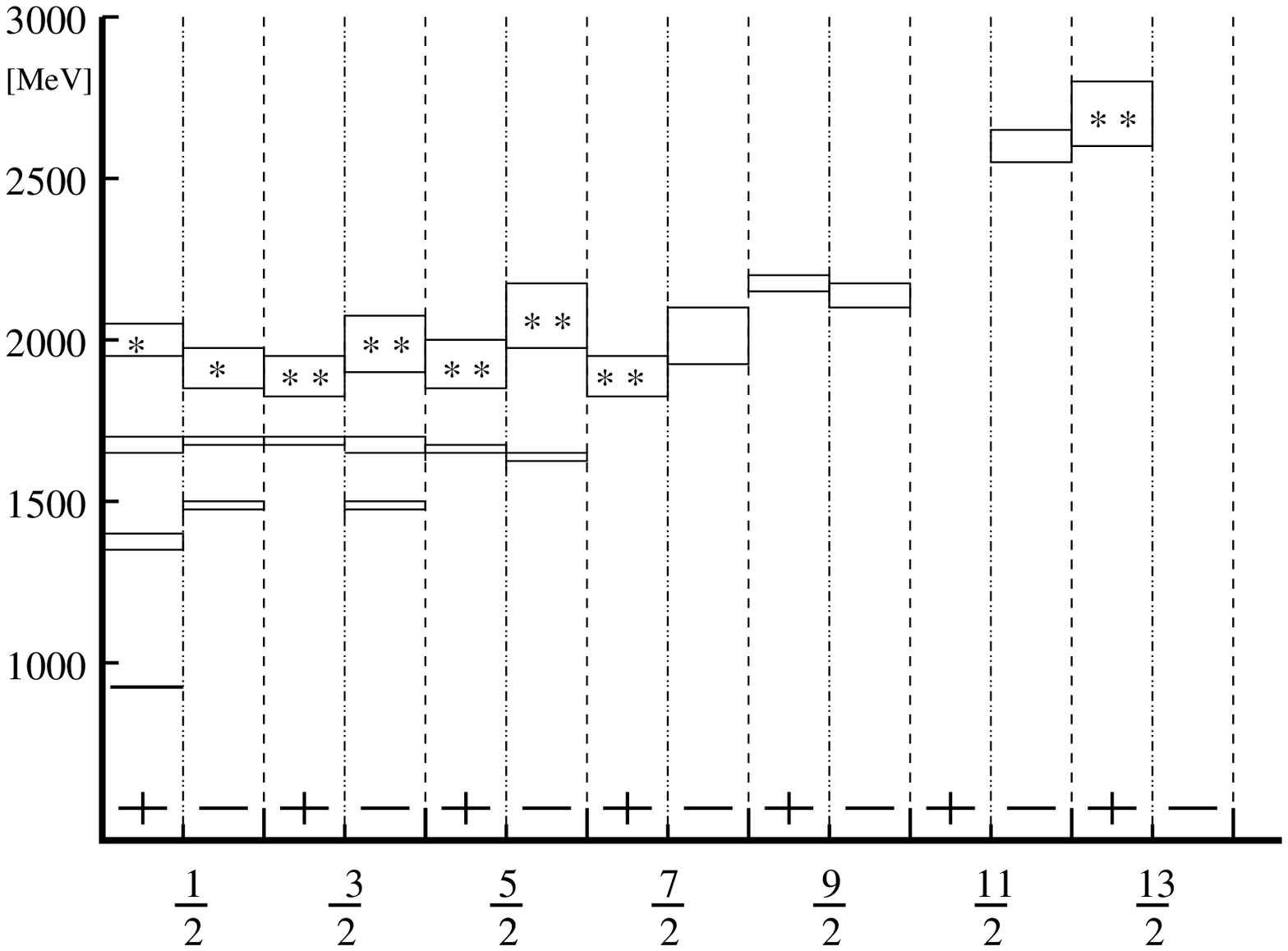}
\includegraphics*[width=5cm,angle=-90]{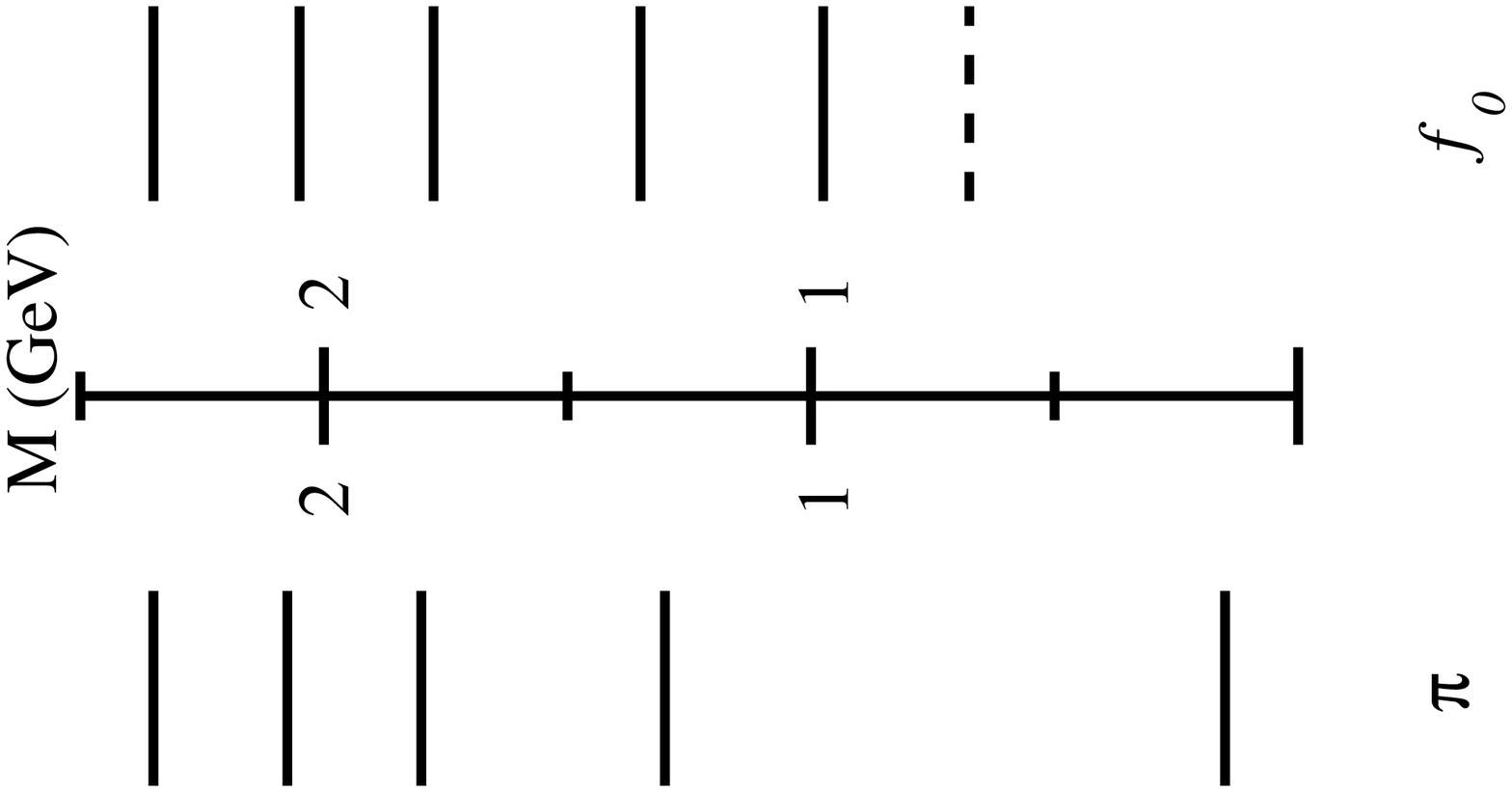}
\end{center}
\caption{Left panel: excitation spectrum of the nucleon (those
resonances which are not yet established are marked by two or one
stars according to the PDG classification). Right panel: pion and
$n \bar n$ $f_0$ spectra. }
\end{figure}

\section{A few words about chiral symmetry breaking and
low-lying hadrons}

A key to the understanding of the low-lying hadrons is the
spontaneous breaking of chiral symmetry (SBCS). Any interquark
interaction in QCD mediated by the intermediate gluon field, in
the local approximation and once antisymmetrization of the quark
fields has been performed, contains as a part a chiral-invariant
4-fermion interaction \cite{NJL}, $ (\bar \psi \psi)^2 + (\bar
\psi i \gamma_5  \vec \tau \psi)^2. $ The first term represents a
Lorentz-scalar interaction. This interaction is an attraction
between the left-handed quarks and the right-handed antiquarks and
vice versa. When it is treated nonperturbatively in the mean-field
approximation, which is well justified in the vacuum state, it
leads to the condensation of the chiral pairs in the vacuum state,
$ \langle 0 | \bar \psi \psi | 0 \rangle = \langle 0 | \bar \psi_L
\psi_R +  \bar \psi_R \psi_L | 0 \rangle  \neq 0 . $ Hence it
breaks chiral symmetry, which is a nonperturbative phenomenon.
This attractive interaction between bare quarks can be  absorbed
into a mass of a quasiparticle. Each quasiparticle is a coherent
superposition of bare quarks and antiquarks. Bare particles have
both well-defined helicity and chirality, while quasiparticles
have only definite helicity and contain a mixture of bare quarks
and antiquarks with opposite chirality.
 These quasiparticles with dynamical mass can
be associated with the constituent quarks. An important feature
is that this dynamical mass appears only at low momenta, below
the ultraviolet cutoff  $\Lambda$ in the NJL model, i.e. where
the low-momentum attractive interaction between quarks is
operative. All quarks with momenta higher than $\Lambda$ remain
undressed. In reality, of course, this step-function behaviour
of the dynamical mass should be substituted by some smooth function.

\begin{figure}
\begin{center}
\includegraphics*[width=5cm]{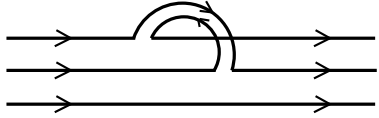}
\end{center}
\caption{Pion-exchange between valence quarks in the low-lying
baryons.}
\end{figure}

Once the chiral symmetry is spontaneously broken, then there must
appear collective massless Goldstone excitations. Microscopically
their zero mass is provided by the  term $(\bar \psi i \gamma_5
\vec \tau \psi)^2$. This term represents an attraction between the
constituent quark and the antiquark with the pion quantum numbers.
Without this term the pion would have a mass of $2M$. When this
term is nonperturbatively and relativistically iterated the
attraction between the constituent quarks in the pion exactly
compensates the $2M$ energy and the pion becomes massless. This
happens because of the underlying chiral symmetry since it is this
symmetry which dictates that the strengths of the interactions
represented by the first and by the second terms in the NJL
Hamiltonian are equal. So the pion is a relativistic bound state
of two quasiparticles. It contains $\bar Q Q, \bar Q Q \bar Q Q,
...$ Fock components. The pion (as any Goldstone boson) is a
highly collective excitation in terms of the original (bare)
quarks and antiquarks  $q$ and $\bar q$ because the quasiparticles
$Q$ and $\bar Q$  themselves are coherent collective excitations
of bare quarks.

Now we will go to the low-lying baryons. A basic ingredient of the
chiral quark picture of Manohar and Georgi \cite{MG} is that the
constituent quarks inside the nucleon are strongly coupled to the
pion field and this coupling is regulated by the
Goldberger-Treiman relation. Why this must be so can be seen
directly from the Nambu and Jona-Lasinio mechanism of chiral
symmetry breaking. Then in the low-momentum regime (which is
responsible for masses) the {\it low-lying} baryons in the light
flavor sector can be approximated as systems of three confined
constituent quarks with the residual interaction mediated by the
Goldtone boson field \cite{GR}. Such a model was designed to solve
a problem of the low-lying baryon spectroscopy. Microscopically
this residual interaction appears from the t-channel iterations of
those gluonic interactions in QCD which are responsible for chiral
symmetry breaking \cite{GV}, see Fig. 2. An essential feature of
this residual interaction is that it is a flavor- and
spin-exchange interaction and contributes to the baryon energy at
the order $N_c$ while the contribution of the color-magnetic
interaction appears only at $1/N_c$. This specific form of the
residual interaction between valence
 constituent quarks in baryons allows us  not only to generate
octet-decuplet splittings but what is  important to solve
at the same time
the long-standing puzzle of the ordering of the lowest excitations
of positive and negative parity in the $u,d,s$ sector. This physics
is  a subject of intensive lattice studies and recent
results \cite{KENTUKKY,SASAKI,BGR,SIMULA} do show that the correct
ordering can be achieved only close to the chiral limit and hence
is related to spontaneous breaking of chiral symmetry.
The  results  \cite{BGR} also
evidence a node in the wave function of the
 radial excitation  of the nucleon (Roper resonance)
 which is consistent with the 3Q leading Fock component
of this state. It is also important to realize for the following
discussion, that all these effects of the pion cloud  are effects
of {\it quantum fluctuations of the quark fields}, which is well
seen from Fig. 2.

This physics and effective degrees of freedom, which are based on
the spontaneous breaking of chiral symmetry, are relevant,
however, only to the low-lying hadrons. In the high-lying hadrons
chiral symmetry is restored.

\section{Chiral symmetry restoration in excited hadrons by definition}

 By definition chiral
 symmetry restoration means the following. In QCD
hadrons with quantum numbers $\alpha$ are created when one applies
the local interpolating field (current) $J_\alpha$ with such
quantum numbers on the vacuum $|0\rangle$. Then all
 hadrons that are created by the given interpolator appear
as intermediate states in the two-point correlator,

\begin{equation}
\Pi =\imath \int d^4x ~e^{\imath q x} \langle 0 | T \{ J_\alpha
(x) J^\dagger_\alpha (0) \} |0\rangle, \label{corr}
\end{equation}

\noindent where all possible Lorentz and Dirac indices (specific
for a given interpolating field) have been omitted. Consider two
local interpolating fields $J_1(x)$ and $J_2(x)$ which are
connected by a chiral transformation (or by a $U(1)_A$
transformation), $ J_1(x) = UJ_2(x)U^\dagger. $
 Then, if the vacuum
was invariant under the chiral group, $U|0\rangle = |0\rangle,$ it
follows from (\ref{corr}) that the spectra created by the
operators $J_1(x)$ and $J_2(x)$ would be identical. We know that
in QCD one finds $U|0\rangle \neq |0\rangle.$ As a consequence the
spectra of the two operators must be in general different.
However, it happens that the noninvariance of the vacuum becomes
unimportant (irrelevant) high in the spectrum. Then the spectra of
both operators become close at large masses and asymptotically
identical. This  means that chiral symmetry is effectively
restored. We stress that this effective chiral symmetry
restoration does not mean that chiral symmetry breaking in the
vacuum disappears, but that the role of the quark condensates that
break chiral symmetry in the vacuum becomes progressively less
important high in the spectrum. One could say, that the valence
quarks in high-lying hadrons {\it decouple} from the QCD vacuum.
In order to avoid a confusion with the chiral symmetry restoration
in the vacuum state at high temperature or density one also refers
this phenomenon as {\it chiral symmetry restoration of the second
kind}.

\section{The quark-hadron duality and chiral symmetry restoration}

A question arises to which extent the chiral symmetry restoration
of the second kind can be theoretically predicted in QCD. There is
a heuristic argument that supports this idea \cite{CG1}. The
argument is based on the well controlled behaviour of the
two-point function (\ref{corr}) at the large space-like momenta
$Q^2 =-q^2$, where the operator product expansion (OPE) is valid
and where all nonperturbative effects can be absorbed into
condensates of different dimensions \cite{SVZ}. The key point is
that all nonperturbative effects of the spontaneous breaking of
chiral symmetry at large $Q^2$ are absorbed into the quark
condensate $\langle \bar q q \rangle$ and other quark condensates
of higher dimension. However, the contribution of these
condensates to the correlation function is regulated by the Wilson
coefficients. The latter ones are proportional to $(1/Q^2)^n$,
where the index $n$ is determined by the quantum numbers of the
current $J$ and by the dimension of the given quark condensate.
Hence, at large enough $Q^2$ the two-point correlator becomes
approximately chirally symmetric. At these high $Q^2$ a matching
with the perturbative QCD (where no SBCS occurs) can be done. In
other words, though chiral symmetry is broken in the vacuum and
all chiral noninvariant condensates are not zero, their influence
on the correlator at asymptotically high $Q^2$ vanishes. This is
in contrast to the situation of low values of $Q^2$, where the
role of chiral symmetry breaking in the vacuum is crucial. Hence,
at $Q^2 \rightarrow \infty$ one has

\begin{equation}
\Pi_{J_1}(Q^2) - \Pi_{J_2}(Q^2) \sim \frac{1}{Q^n}, \; \; n>0 \; .
\label{highQ}
\end{equation}

Now we can use the causality of the local field theory and hence
the analyticity of the two-point function. Then we can invoke into
analysis a dispersion relation,

\begin{equation}
\Pi_J(Q^2) \, = \,  \int {\rm d}s \frac{\rho_J(s)}{Q^2 + s - i \epsilon},~~~~
\rho_J (s) \equiv \frac{1}{\pi} \rm{Im} \left ( \Pi_j(s) \right ).
\label{kl}
\end{equation}

\noindent
Since the large $Q^2$ asymptotics of the correlator
is given by the leading term of
the  perturbation theory,
then the asymptotics of $\rho(s)$ at $s \rightarrow \infty$ must
also be given by the same term of the perturbation theory if the
spectral density approaches a constant value (if it oscillates, then
it must oscillate around the perturbation theory value). Hence
both spectral densities $\rho_{J_1}(s)$ and $\rho_{J_2}(s)$ at
$s \rightarrow \infty$ must approach the same value and the spectral
function becomes chirally symmetric. This theoretical expectation,
that the high $s$ asymptotics of the spectral function is well described
by the leading term of the perturbation theory has been tested e.g.
in the process $e^+e^- \rightarrow hadrons$ , where the interpolator
is given by the usual electromagnetic vector current.
 Similarly, the vector and the axial vector
spectral densities must coincide in the chiral symmetry restored
regime. They have been measured in the $\tau$ decay by the ALEPH
and OPAL collaborations at CERN \cite{ALEPH,OPAL}. It is well seen
from the results that while the difference between both spectral
densities is large at the masses of the $\rho(770)$ and
$a_1(1260)$, it becomes
 strongly reduced towards
$m=\sqrt s \sim 1.7$ GeV.

The question arises then what is the functional behaviour that
determines approaching the chiral-invariant regime at large $s$?
Naively one would expect that the operator product expansion
of the two-point correlator, which is valid in the deep
Euclidean domain, could help us. This is not
so, however, for two reasons. First of all, we know
phenomenologically only the lowest dimension quark condensate.
Even though this condensate dominates as a chiral symmetry
breaking measure at the very large space-like $Q^2$, at
smaller $Q^2$ the higher dimensional condensates, which
are suppressed by inverse powers of $Q^2$, are also
important. These condensates are not known, unfortunately.
But even if we knew all  quark condensates
up to a rather high dimension, it would not help us. This is
because the OPE is only an asymptotic expansion \cite{Z}.
While such kind of expansion is very useful in the space-like
region, it does not define any analytical solution which could
be continued to the time-like region at finite $s$. While
convergence of the OPE can be improved by means of the Borel
transform and it makes it useful for SVZ sum rules for the
low-lying hadrons, this cannot be done for the higher states.
So in order to estimate chiral symmetry restoration effects
 one indeed needs a microscopic theory that would incorporate
{\it at the same time} chiral symmetry breaking and confinement.

\section{Chiral and $U(1)_A$ restorations as a manifestation
of the semiclassical regime}

While the argument above on the asymptotic symmetry properties of
spectral functions is rather robust (it is based actually only on
the asymptotic freedom of QCD at large space-like momenta and on
the analyticity of the two-point correlator), {\it a-priori} it is
not clear whether it can be applied to bound state systems, which
the hadrons are. Indeed, it can happen that the asymptotic
symmetry restoration applies only to that part of the spectrum,
which is above the resonance region
 (i.e. where the current creates jets but not isolated hadrons). So
the question arises whether it is possible to prove (or at
least justify) the symmetry restoration in highly excited
{\it isolated} hadrons. It is shown in ref. \cite{G5}
that both chiral
and $U(1)_A$ restorations in highly excited isolated hadrons  can
be anticipated as a direct consequence of the semiclassical
regime in the highly excited hadrons, indeed.

At large $n$ (radial quantum number) or at large angular
momentum $J$ we know that in quantum systems the {\it semiclassical}
approximation (WKB) {\it must} work. Physically this approximation
applies in these cases because the de Broglie wavelength of
particles in the system is small in comparison with the
scale that characterizes the given problem. In such a system
as a hadron the scale is given by the hadron size while the
wavelength of valence quarks is given by their momenta. Once
we go high in the spectrum the size of hadrons increases as well as
 the typical momentum of valence quarks.
This is why a highly excited hadron  can be described semiclassically
in terms of the underlying quark  degrees of freedom.

A physical content of the semiclassical approximation is most
transparently given by the path integral. The contribution of the
given path to the path integral, $ \sim e^{iS(q)/\hbar}, $ is
regulated by the action $S(q)$ along the path $q(x,t)$. The
semiclassical approximation  applies when the action in the system
$S \gg \hbar$. In this case the whole amplitude (path integral) is
dominated by the classical path $q_{cl}$ (stationary point) and
those paths that are infinitesimally close to the classical path.
All other paths that differ from the classical one by an
appreciable amount  do not contribute. These latter paths would
represent the quantum fluctuation effects. In other words, in the
semiclassical case the quantum fluctuations effects are strongly
suppressed and vanish asymptotically. The classical path is a
tree-level contribution to the path integral
 and keeps all symmetries of the classical
Lagrangian. Its contribution is of the order $S/\hbar$. The
quantum fluctuations of the fields contribute at the orders
$(\hbar/S)^0$ (one loop), $(\hbar/S)^1$ (two loops), etc.

The $U(1)_A$ as well as spontaneous $SU(2)_L \times SU(2)_R$
breakings  result from quantum fluctuations of the quark fields.
However, in a quantum system with large enough $n$ or  $J$ the
quantum fluctuations contributions must be relatively suppressed
and vanish asymptotically. Then it follows that in such systems
both the chiral and $U(1)_A$ symmetries must be restored. Hence at
large hadron masses (i.e. with either large $n$ or large $J$) we
must observe the symmetries of the classical QCD Lagrangian. This
is precisely what we see phenomenologically. In the nucleon
spectrum the doubling appears either at large $n$ excitations of
baryons with the given small spin or in resonances of large spin.
Similar features persist in the delta spectrum. In the meson
spectrum the doubling is obvious for large $n$ excitations of
small spin mesons and there are signs of doubling of large spin
mesons (the data are, however, sparse). It would be certainly
interesting and important to observe systematically multiplets of
parity-chiral and parity-$U(1)_A$ groups  (or, sometimes, when the
chiral and $U(1)_A$ transformations connect {\it different}
hadrons, the multiplets of the $U(2)_L \times U(2)_R$ group
\cite{G2} ).

While the argument above is  robust, theoretically it is
not clear {\it a-priori} whether isolated hadrons still
exist at excitation energies where a semiclassical regime
is achieved. Hence it is conceptually important to demonstrate that
in QCD hadrons still exist, while the dynamics inside such
 hadrons already is semiclassical. We do not know how to prove it for
$N_c =3$. However, the large $N_c$ limit of QCD \cite{H}, while
keeping all basic properties of QCD like asymp\-totic freedom and
chiral symmetry, allows for a significant simplification. In this
limit it is known that all mesons represent  narrow states, i.e.
they are stable against strong decays. At the same time the
spectrum of mesons is infinite (see, e.g., \cite{W}). The latter
is necessary to match the two-point function in the perturbation
theory regime (which contains logarithm) at large space-like
momenta with the discrete spectral sum in the dispersion integral.
Then one can always excite a meson of any
 arbitrary large energy, which is of any arbitrary large size.
 In such a meson $S \gg \hbar$. Hence a description
 of this meson necessarily must  be semiclassical. Then the equation
 of motion in such a meson must be according to some yet
 unknown solution of the classical QCD Lagrangian
for a colorless meson. Hence
 it must exhibit chiral and $U(1)_A$ symmetries.

Actually we do not need the exact $N_c = \infty$ limit for this
statement. It can be formulated in the following way. For any
large $S \gg \hbar$ there always exist such $N_c$ that the
isolated meson with such an action does exist and can be described
semiclassically. From the  empirical fact that we observe
multiplets of chiral and $U(1)_A$ groups high in the hadron
spectrum it follows that $N_c=3$ is large enough for this purpose.
 \footnote{That the quantum
fluctuations effects vanish in the quantum bound state systems
at large $n$ or $J$
is well known e.g. from the Lamb shift. The Lamb shift is a result
of the radiative corrections (which represent
effects of quantum fluctuations of electron and electromagnetic
fields)
 and vanishes  as $1/n^3$,
and also very fast with increasing $J$. As a consequence high in the
hydrogen spectrum  the symmetry of the classical Coulomb
potential gets restored.
The other well-known example which clearly illustrates the
point is the 't Hooft model \cite{HM} (QCD in 1+1 dimensions).
In this model in the regime $N_c \rightarrow \infty$,
$m_q \rightarrow 0$, $m_q \gg g \sim 1/\sqrt N_c$, the spectrum
of the high-lying states is known exactly, $M_n^2 \sim n$. The
chiral symmetry of the Lagrangian is broken (with no contradiction
with the Coleman theorem since in this specific regime everything
is determined by $N_c = \infty$ , for any large but finite $N_c$
the chiral symmetry is not broken in agreement with the Coleman
theorem ), which is reflected by the fact that the positive
and negative parity states are not degenerate and alternate
in the spectrum. However, the mass difference between the
 neighbouring positive and negative parity states is
$m_+ - m_- \sim 1/\sqrt n$ and  vanishes high in the spectrum
since the effect of quantum fluctuations dies out
high in the spectrum.
}

\begin{figure}
\begin{center}
\includegraphics*[width=7cm]{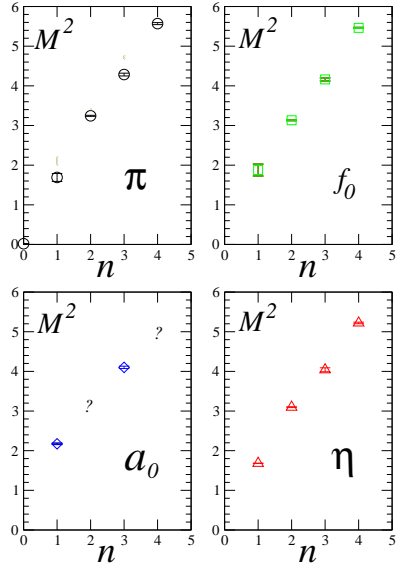}
\end{center}
\caption{Radial Regge trajectories for the four successive
high-lying $J=0$ mesons.}
\end{figure}

\section{Chiral multiplets of excited mesons}

A detailed classification of the chiral and $U(1)_A$ multiplets
for excited mesons, based on the recent experimental data obtained
from proton-antiproton annihilation at LEAR \cite{BUGG1,BUGG2}, is
given in ref. \cite{G2}. Here we limit ourselves only to a few
typical and spectacular examples.

 Consider first the mesons of spin $J=0$,
which are the $ \pi,f_0,a_0$ and $\eta$ mesons with $u,d$ quark
content only. If one looks at the upper part of the meson
spectrum, then one notices that the four successive  excited $\pi$
mesons and the corresponding $\bar n n$ $f_0$ mesons form
approximate chiral pairs and fill out (1/2,1/2) representations of
the chiral group. This is well seen from Fig. 1. This pattern is a
clear manifestation from chiral symmetry restoration.

A similar behaviour is observed from a comparison of the $a_0$ and
$\eta$ masses. However, there are two missing $a_0$ mesons which
must be discovered in order to complete all chiral multiplets.
There is little doubt that these missing $a_0$ mesons do exist. If
one puts the four high-lying $\pi$, $\bar n n$ $f_0$, $a_0$ and
$\bar n n$ $\eta$ mesons on the {\it radial} Regge trajectories,
see Fig. 3, one clearly notices that the two missing $a_0$ mesons
lie on the linear trajectory with the same slope as all other
mesons. If one reconstructs these missing $a_0$ mesons according
to this slope, then a pattern of the $a_0 - \eta$ chiral partners
appears, similar to the one for the $\pi$ and $f_0$ mesons.

For the $J \geq 1$ mesons the classification is a bit
more complicated.
 Below we show the chiral patterns for the $J=2$
mesons, where the data set seems to be complete (masses are in
MeV).

\begin{center}{\bf (0,0)}\\

{$\omega_2(0,2^{--})~~~~~~~~~~~~~~~~~~~~~~~f_2(0,2^{++})$}\\
\medskip
{$1975 \pm 20   ~~~~~~~~~~~~~~~~~~~~~~~~1934 \pm 20$}\\
{$2195 \pm 30   ~~~~~~~~~~~~~~~~~~~~~~~~2240 \pm 15$}\\

\bigskip
{\bf (1/2,1/2)}\\

{$\pi_2(1,2^{-+})~~~~~~~~~~~~~~~~~~~~~~~f_2(0,2^{++})$}\\
\medskip
{$2005 \pm 15   ~~~~~~~~~~~~~~~~~~~~~~~~2001 \pm 10$}\\
{$2245 \pm 60   ~~~~~~~~~~~~~~~~~~~~~~~~2293 \pm 13$}\\

\bigskip
{\bf (1/2,1/2)}\\

{$a_2(1,2^{++})~~~~~~~~~~~~~~~~~~~~~~~ \eta_2(0,2^{-+})$}\\
\medskip
{ $2030 \pm 20  ~~~~~~~~~~~~~~~~~~~~~~~~2030 ~\pm ~?$}\\
{ $2255 \pm 20 ~~~~~~~~~~~~~~~~~~~~~~~~2267 \pm 14$}\\

\bigskip
{\bf (0,1)+(1,0)}\\

{$a_2(1,2^{++})~~~~~~~~~~~~~~~~~~~~~~~\rho_2(1,2^{--})$}\\
\medskip
{ $1950^{+30}_{-70}~~~~~~~~~~~~~~~~~~~~~~~~~~~1940 \pm 40$}\\
{ $2175 \pm 40  ~~~~~~~~~~~~~~~~~~~~~~~~2225 \pm 35$}\\
\end{center}

\noindent We see systematic patterns of chiral symmetry
restoration. In particular, the amount of $f_2(0,2^{++})$ mesons
coincides with the combined amount of $\omega_2(0,2^{--})$ and
$\pi_2(1,2^{-+})$ states. Similarly, the number of $a_2(1,2^{++})$
states is the same as the number of $\eta_2(0,2^{-+})$ and
$\rho_2(1,2^{--})$ together. All chiral multiplets are complete.
While the masses of some of the states can and will be  corrected
in future experiments, if  new  states might be discovered in this
energy region in other types of experiments, they should be either
$\bar s s$ states or glueballs.

It is important to see whether there are also signatures
of the $U(1)_A$ restoration.
  Those interpolators (states) that are members
of the $(0,0)$ and $(0,1)+(1,0)$ representations of
$SU(2)_L \times SU(2)_R$ are invariant with respect to
$U(1)_A$. However,  interpolators (states) from the {\it distinct}
(1/2,1/2) representations which have the same isospin but
opposite parity  transform into each other under $U(1)_A$.
If the corresponding states are systematically (approximately)
degenerate, then it is a signal that $U(1)_A$ is restored.
 In what follows we show
that it is indeed the case.

\begin{center}

{$f_2(0,2^{++})~~~~~~~~~~~~~~~~~~~~~~~\eta_2(0,2^{-+})$}\\
\medskip
{$2001 \pm 10  ~~~~~~~~~~~~~~~~~~~~~~~~2030 ~\pm ~?$}\\
{$2293 \pm 13   ~~~~~~~~~~~~~~~~~~~~~~~~2267 \pm 14$}\\

\bigskip

{$\pi_2(1,2^{-+})~~~~~~~~~~~~~~~~~~~~~~~a_2 (1,2^{++})$}\\
\medskip
{$2005 \pm 15  ~~~~~~~~~~~~~~~~~~~~~~~~ 2030 \pm 20$}\\
{$2245 \pm 60 ~~~~~~~~~~~~~~~~~~~~~~~~ 2255 \pm 20 $}\\
\end{center}

 We see clear approximate doublets of $U(1)_A$ restoration. Hence
two distinct (1/2,1/2) multiplets of $SU(2)_L \times SU(2)_R$ can
be combined into one multiplet of $U(2)_L \times U(2)_R$. So we
conclude that the whole chiral symmetry  of the QCD Lagrangian
$U(2)_L \times U(2)_R$ gets appro\-ximately restored high in the
hadron spectrum.

It is useful to quantify the effect of chiral symmetry breaking
(restoration). An obvious parameter that characterizes effects of
chiral symmetry breaking is a relative mass splitting within the
chiral multiplet. Let us define the {\it chiral asymmetry} as

\begin{equation}
\chi = \frac{|M_1 - M_2|}{(M_1+M_2)},
\label{chir}
\end{equation}

\noindent
where $M_1$ and $M_2$ are masses of particles within the
same multiplet. This parameter gives a quantitative measure
of chiral symmetry breaking at the leading (linear) order
and has the interpretation of the part
 of the hadron mass  due to chiral
symmetry breaking. For the low-lying states the chiral asymmetry
is typically 0.2 - 0.6 which can be seen e.g. from a comparison of
the $\rho(770)$ and $a_1(1260)$ or the  $\rho(770)$ and
$h_1(1170)$ masses. If the chiral asymmetry is large as above,
then it makes no sense to assign a given hadron to the chiral
multiplet since its wave function is a strong mixture of different
representations and we have to expect also large {\it nonlinear}
symmetry breaking effects. However, at meson masses of about 2 GeV
the chiral asymmetry is typically within 0.01 and in this case the
hadrons can be believed to be  members of  multiplets with a tiny
admixture of other representations.

\section{Chiral multiplets of excited baryons}

Now we will consider chiral multiplets of excited baryons \cite{CG1}.
 The only possible representations of the $SU(2)_L \times SU(2)_R$
 group that are invariant under parity are
$(1/2,0) \oplus (0,1/2)$, $(1/2,1) \oplus (1,1/2)$
 and $(3/2,0) \oplus (0,3/2)$. Since chiral symmetry and
parity do not constrain the possible spins of the states
these multiplets can correspond to states of any fixed spin.

 A phenomenological consequence of the effective
restoration of chiral symmetry high in $N$ and $\Delta$ spectra is
that the baryon states will fill out  the irreducible
representations of the parity-chiral group. If $(1/2,0) \oplus
(0,1/2)$ and $(3/2,0) \oplus (0,3/2)$ multiplets were realized in
nature, then the spectra of highly excited nucleons and deltas
would consist of parity doublets. However, the energy of the
parity doublet with  given spin in the nucleon spectrum {\it
a-priori} would not be degenerate with the the doublet with the
same spin in the delta spectrum; these doublets would belong to
different representations, {\it i.e.} to distinct multiplets and
their energies are not related.   On the other hand, if $(1/2,1)
\oplus (1,1/2)$ were realized, then the high-lying states in the
$N$ and $\Delta$ spectra would have a $N$ parity doublet and a
$\Delta$ parity doublet with the same spin and which are
degenerate in mass. In either case the high-lying spectrum must
systematically consist of parity doublets.

If one looks carefully at the nucleon spectrum, see Fig. 1, and
the delta spectrum one notices that the systematic parity doubling
in the nucleon spectrum appears at masses of 1.7 GeV and above,
while the parity doublets in the delta spectrum insist at  masses
of 1.9 GeV and higher. This means that the parity doubling in both
cases is seen at approximately the same excitation energy with
respect to the corresponding ground state. This is because not an
absolute value of the energy is important
 in order to approach
 the chiral symmetry restoration regime  but rather $n$ or $J$. In both
nucleon and delta spectra parity doubling persists at the same
values of $n$ and $J$.
 This fact implies that at
least those nucleon doublets that are seen at $\sim 1.7$ GeV
belong to the $(1/2,0) \oplus (0,1/2)$ representation. If
approximate mass degeneracy between some of the $N$ and $\Delta$
doublets at $M \geq 1.9$ GeV is accidental, then the baryons in
this mass region are organized according to $(1/2,0) \oplus
(0,1/2)$ for $N$ and $(3/2,0) \oplus (0,3/2)$ for $\Delta$
parity-chiral doublets. If not, then some of the high-lying
doublets form $(1/2,1) \oplus (1,1/2)$ multiplets. It can also be
possible that in the narrow energy interval more than one parity
doublet in the nucleon and delta spectra is found for a given
spin. This would then mean that different doublets  belong to
different parity-chiral multiplets. Systematic experimental
exploration of the high-lying states is required in order to
assign unambiguously baryons to the multiplets.

\section{Chiral symmetry restoration and the string (flux tube) picture}

Before discussing a model for highly excited hadrons that is
compatible with the chiral symmetry restoration and parity
doubling it is useful to answer the question whether the
nonre\-lativistic or relativized potential models  based on the
$^{2S+1}L_J$ description like the traditional constituent quark
model can explain it. This question can be definitely answered
"no"; for a transparent explanation see the review \cite{G3}.

We have already discussed that in highly excited hadrons the
valence quark motion has to be described semiclassically and at
the same time their chirality (helicity) must be fixed. Also the
gluonic field should be described semiclassically. All this gives
an increasing support for a string picture \cite{NAMBU,NO} of
highly excited hadrons. Indeed, if one assumes that the quarks at
the ends of the string have definite chirality, see Fig. 4, then
all hadrons will appear necessarily in chiral and $U(1)_A$
multiplets \cite{G4}. The latter picture is very natural and is
well compatible with the Nambu string picture. The ends of the
string in the Nambu model move with the velocity of light. Then,
(it is an extension of the Nambu model) the quarks at the ends of
the string must have definite chirality. In this way one is able
to explain at the same time both
Regge trajectories and parity doubling.\\

\begin{figure}
\begin{center}
\includegraphics*[width=3cm,angle=-90]{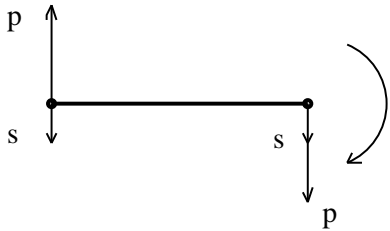}
\end{center}
\caption{ Rotating string with the right and the left quarks at the ends.}
\end{figure}

There is one additional and important byproduct of this
picture \cite{G4}.
The spin-orbit operator $\vec \sigma \cdot \vec L$ does not
commute with the helicity operator $\vec \sigma \cdot \vec \nabla$.
Hence the spin-orbit interaction of quarks with the fixed
chirality or helicity is absent. In particular, this is also true
for the spin-orbit force due to the Thomas precession

\begin{equation}
U_T = -\vec \sigma \cdot \vec \omega_T \sim \vec \sigma
 \cdot [\vec v, \vec a]
\sim \vec v \cdot [\vec v, \vec a] =0,
\label{T}
\end{equation}

\noindent
where $U_T$ is the energy of the interaction and
$\vec \omega_T$, $\vec v$ and $\vec a$ are the angular frequency
of Thomas precession, velocity of the quark and its acceleration,
 respectively.

The absence of the spin-orbit force in the chirally restored
regime is a very welcome feature because it is a well-known
empirical fact that the spin-orbit force is either vanishing or
very small in the spectroscopy in the $u,d$ sector. In addition,
for the rotating string $ \vec \sigma (i) \cdot \vec R(i)=0, $ $
 \vec \sigma (i) \cdot \vec R(j)=0.
 $
 The relations above
immediately imply that the possible tensor interactions
of quarks related to the string dynamics should
be absent, once the chiral symmetry is restored.

\section{Conclusions}

We have demonstrated  that the chiral symmetry of QCD is crucially
important to understand the physics of hadrons in the $u,d$ (and
possibly in the $u,d,s$) sector. The low-lying hadrons are mostly
driven by the spontaneous breaking of chiral symmetry. This
breaking determines the physics and effective degrees of freedom
in the low-lying hadrons. However, this physics is relevant only
to the low-lying hadrons. In the high-lying hadrons chiral
symmetry is restored, which is referred to as effective chiral
symmetry restoration or chiral symmetry restoration of the second
kind. A direct manifestation of the latter phenomenon is a
systematic appearance of the approximate chiral multiplets of the
high-lying hadrons. The essence of the present phenomenon is that
the quark condensates which break chiral symmetry in the vacuum
state (and hence in the low-lying excitations over the vacuum)
become simply irrelevant (unimportant) for the physics of the
highly excited states. The physics here is as if there was no
chiral symmetry breaking in the vacuum. The valence quarks
decouple from the quark condensates and consequently the notion of
the constituent quarks with dynamical mass induced by chiral
symmetry breaking becomes inadequate in the highly excited
hadrons. This happens because in the highly excited hadrons the
effects of quantum fluctuations vanish and a semiclassical regime
is manifest.

Hence  physics of the high-lying hadrons is mostly physics
of confinement acting between the light quarks. Their very small current
mass strongly distinguishes this physics from the
physics of the heavy quarkonium, where chiral symmetry is irrelevant
 and the string (flux tube) can be approximated as a static potential
acting between the slowly moving heavy quarks. In the light
hadrons in contrast the valence quarks are ultrarelativistic and their
fermion
nature requires them to have a definite chirality. Hence the
high-lying hadrons in the $u,d$ sector  open a door to the regime of
dynamical strings with chiral quarks at the ends. Clearly a
systematic experimental exploration of the high-lying hadrons is
required which is an interesting and important task and which should
be of highest priority at the existing and forthcoming accelerators.\\

\end{document}